\documentclass[usenatbib]{mn2e}

\bibpunct{(}{)}{;}{a}{}{,}

\newcommand{\ltsima}{$\buildrel < \over \sim$}
\newcommand{\lsim}{\lower.5ex\hbox{\ltsima}}
\newcommand{\gtsima}{$\buildrel > \over \sim$}
\newcommand{\gsim}{\lower.5ex\hbox{\gtsima}}

\newcommand{\aap}{A\&A}

\usepackage{gensymb}
\usepackage{graphicx}
\usepackage{amssymb}
\usepackage{epsfig}
\usepackage{natbib}
\usepackage{epsf,palatino,url,graphicx,wrapfig,array,setspace,amssymb,fancyhdr,multirow,lscape,appendix,rotating,amsmath,xcolor}

\def\ion#1#2{#1$\;${\small\rm\@Roman{#2}}\relax}

\title[BeX cat ]{Catalogue of Be/X-ray binary systems in the Small Magellanic Cloud: X-ray, optical \& IR properties}

\author[M.J. Coe]{{M. J.~Coe \& J. Kirk} \\
Physics and Astronomy, University of Southampton, SO17
1BJ, UK. \\
}

\begin{document}

\date{June 3 2015}


\maketitle

\label{firstpage}

\begin{abstract}

{
This is a catalogue of $\sim$70 X-ray emitting binary systems in the Small Magellanic Cloud (SMC) that contain a Be star as the mass donor in the system and a clear X-ray pulse signature from a neutron star. The systems are generally referred to as Be/X-ray binaries. It lists all their known binary characteristics (orbital period, eccentricity), the measured spin period of the compact object, plus the characteristics of the Be star (spectral type, size of the circumstellar disk, evidence for NRP behaviour). For the first time data from the Spitzer Observatory are combined with ground-based data to provide a view of these systems out into the far-IR.

Many of the observational parameters are presented as statistical distributions and compared to other similar similar populations (eg isolated Be \& B stars) in the SMC, and to other Be/X-ray systems in the Milky Way. In addition previous important results are re-investigated using this excellently homogenous sample. In particular, the evidence for a bi-modality in the spin period distribution is shown to be even stronger than first proposed, and the correlation between orbital period and circumstellar disk size seen in galactic sources is shown to be clearly present in the SMC systems and quantised for the first time.

}

\end{abstract}

\begin{keywords}
stars:neutron - X-rays:binaries
\end{keywords}

\section{Introduction and background}

The Be/X-ray binary systems represent the largest sub-class of all High Mass X-ray Binaries (HMXB). A survey of the literature reveals that of the $\sim$240 HMXBs known in our Galaxy and the Magellanic Clouds ~\citep{Liu2005}, more than 50\%
fall within this class of binary. In fact, in recent years a substantial population of HMXBs has emerged in the Small Magellanic Cloud (SMC), comparable in number to the Galactic population, though unlike the Galactic population, all except one of the SMC HMXBs are Be star systems - referred to here as Be/X-ray systems. In these systems the orbit of the Be star and the compact object, presumably a neutron star, is generally wide and eccentric. X-ray outbursts are normally associated with the
passage of the neutron star close to the circumstellar disk ~\citep{NO2001}, and generally are classified as Types I or II ~\citep{SWR1986}. The Type I outbursts occur periodically at the time of the periastron passage of the neutron star, whereas Type II outbursts are much more extensive and occur when the circumstellar material expands to fill most, or all of the orbit. See ~\cite{Reig2011} for a comprehensive review of these systems.

The population of such systems in the SMC is extremely valuable as it both provides a large sample (comparable to that known in the Milky Way), and one that is unimpaired by uncertainties in the distance and interstellar absorption. As such it provides an excellent laboratory to study the behaviour of both partners in these interacting binary systems. In this paper we will exclusively address the properties of these SMC sources, but we also note, in passing, that there is a smaller population in the Large Magellanic Cloud that is currently being identified through new X-ray surveys.

\section{The catalogue}

In this paper we present a summary of all such sources in the Small Magellanic Cloud. To be included in this catalogue the source must have exhibited clear evidence for X-ray pulsations at some stage. It is highly probable that there are many more systems that have yet to show such a feature, but already exist in X-ray surveys (see, for example, McBride et al, in prep.).

Presented in Table~\ref{tab:smc1} and in Table~\ref{tab:smc2} are the X-ray properties of the Be/X-ray systems in the SMC. Remarks pertaining to the contents of the tables are as follows:

\begin{itemize}

\item  For convenience, all of the Magellanic Cloud sources are identified by a short name (column one in all the tables) first proposed by ~\cite{Coe2005a}.  This identity is created simply from the acronym SXP (Small magellanic cloud X-ray Pulsar) followed by the pulse period in seconds to three significant figures. Where helpful, other official source names are listed in the tables. We have not attempted to provide a list of references associated with each source because an up to date literature review may be easily and accurately obtained through Simbad\footnote{http://simbad.u-strasbg.fr/simbad/}.

\item Very few systems have reliable measurements of eccentricity - possibly the most secure measurement is one obtained from the Doppler shifting of the pulse period around an orbit. Those systems that have good values are the shorter orbital period systems and listed in ~\cite{Townsend2011} and the measurements are included here in Table ~\ref{tab:smc1}, plus the more recent one for SXP 5.05 ~\citep{Coe2015}. None of the longer period sources in Table ~\ref{tab:smc2} have known eccentricities at this time.

\item There is a column labelled "Other periods". These refer to periodic behaviour detected in the system that is believed not to be at the binary nor the pulse period. Examples of such characteristics are the presence of Non Radial Pulsations associated with the Be star (see, for example, ~\cite{Bird2012}), or a super-orbital period proposed to arise from a precessing accretion disk ~\citep{Coe2013, Andry2011}.

\end{itemize}

\begin{figure}
\includegraphics[angle=-0,width=80mm]{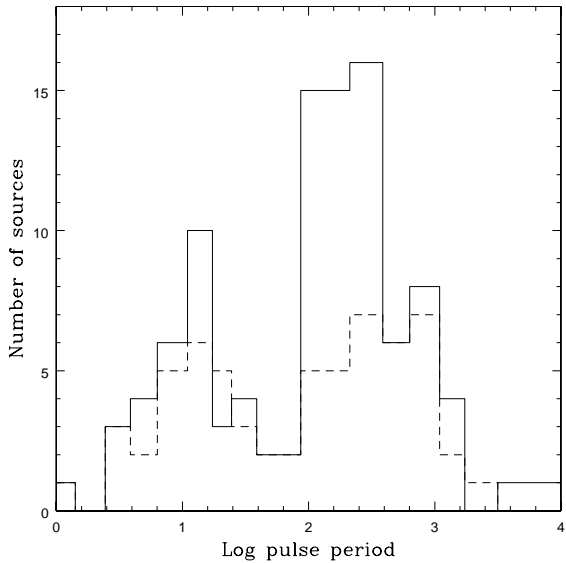}
\caption{The distribution of pulse periods for all the Small Magellanic Cloud sources presented in this work (solid line). It is compared to the same distribution as presented in~\protect\cite{Knigge2011} (dashed line).}
\label{fig:bimod}
\end{figure}

Table~\ref{tab:smcopt} presents the optical and IR characteristics of the mass donor star in the SMC systems. Because this star plays a major role in the observed X-ray characteristics of Be/X-ray systems it is important to include the known data on these components as well. This table contains the following columns:
\begin{itemize}

\item the proposed spectral identification of the Be star (obtained from a wide range of sources, all in published papers, all available through Simbad)

\item the B\&V magnitudes, widely available from several SMC surveys eg ~\cite{Zaritsky2002}

\item the JHK magnitudes from the SMC survey carried out from the Japanese/South African Infra Red Survey Facility (IRSF) and published by ~\cite {Kato2007}

\item the longer wavelength IR photometric values produced from the Spitzer SAGE surveys of the SMC ~\citep{Bolatto2007,Gordon2011,Riebel2015}

\item the measurements of the Equivalent Width (EW) of the H$\alpha$ emission line obtained from several observing runs at both the SALT and SAAO 1.9m telescopes and summarised in ~\cite{Klus2014}.

\end{itemize}

\section{Discussion topics arising from the catalogue}

\subsection{The pulse period bimodality question}

Figure~\ref{fig:bimod} reveals that the increasing number of Be/X-ray systems in the Magellanic Clouds is strengthening the case presented by ~\cite{Knigge2011} that there is a clear bimodality in the pulse period distribution of these systems. Applying the same KMM test discussed in that paper to the updated SMC distribution results in a probability of $1.5\times10^{-4}$ against this being a single distribution. This is compared to $6\times10^{-4}$ for the smaller sample presented in ~\cite{Knigge2011}.

The short and long spin sub-populations contribute about 40\% and 60\% to the total number, respectively, in this updated SMC sample. This is consistent with the ratio determined for the total Magellanic Clouds + Milky Way populations of 35\% to 65\% by ~\cite{Knigge2011}.

Possible explanations for this bimodality currently includes two competing channels for producing supernovae and hence neutron stars ~\citep{Heger2003}, or maybe two modes of accretion ~\citep{Cheng2014}. The former suggestion leads to the possibility of confirmation by predicting different orbital eccentricities for the two populations, but currently the small number of confirmed values makes it difficult to say much at this time. This is one area where better observational results, perhaps through optical RV studies, could play a crucial role in understanding this bimodality phenomenon.

\subsection{IR properties of the SMC sources}

The IR emission in Be/X-ray binary systems shows an excess over the standard stellar spectrum, and is believed to originate from free-free and bound-free emission in the circumstellar disk. It is the material in this disk that is responsible for fuelling the accretion on to the neutron star and hence generating the X-ray behaviour. It is also believed that the presence of the neutron star in orbit around the Be star can constrain the growth and size of the circumstellar disk. One test of this hypothesis is to compare the IR emission from the Be stars in these systems with isolated Be stars, and also also isolated B and B[e] stars.

\begin{figure}
\centering
\includegraphics[angle = -90, width=90mm]{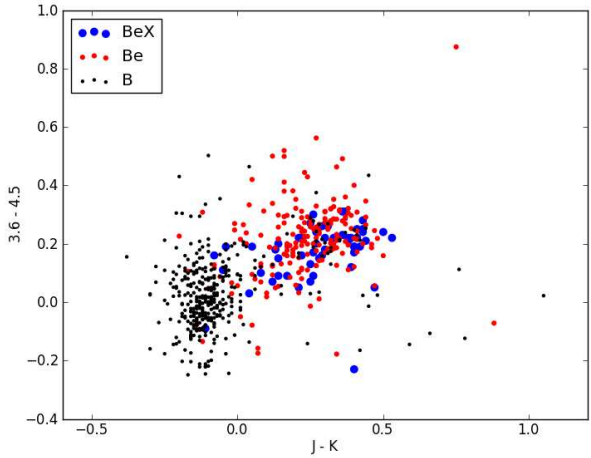}
\caption{Colour-colour comparison plot of Be/X-ray sources, Be stars and B stars populations}
\label{fig:BeXBeBCCD}
\end{figure}

The isolated Be stars from the 2dFS catalogue \citep{Evans2004} were plotted on the same colour-colour diagram (CCD) as the Be/X-ray sources (Figure \ref{fig:BeXBeBCCD}). Both populations had data taken from the Sirius \citep{Kato2007} and the SAGE-SMC catalogues ~\citep{Gordon2011}.  In addition, the B type stars, again from the same 2dFS catalogue, were then added to this plot . This plot showed there to be great similarities in the distributions of the two populations of Be stars, but the disk-free B stars clearly fall elsewhere on this diagram. The B stars can be seen to occupy a region to the left of the CCD and it appears as though the Be and Be/X-ray sources branch off to the right of the main body of B stars. There are some outlying B stars which are distributed amongst the Be and Be/X-ray sources but these may have been incorrectly identified within the respective 2" search radii, or simply been previously misidentified as lacking a circumstellar disk. Since the disks in these systems can sometimes disappear, then Be stars can be thought to be B stars if observed at those times.

\begin{figure}
\centering
\includegraphics[angle = -90,width=60mm]{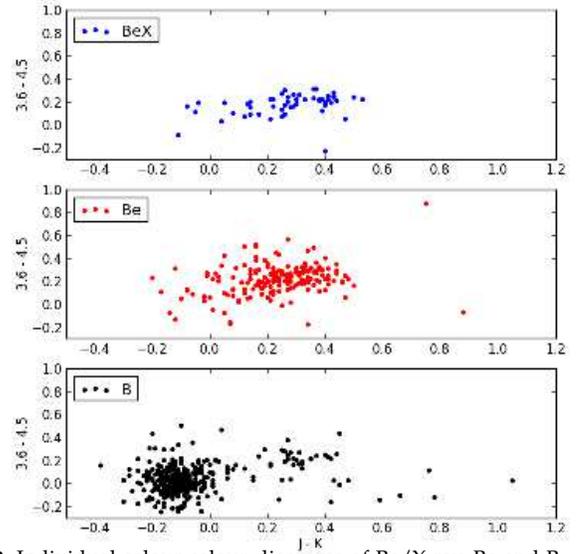}
\caption{Individual colour-colour diagrams of Be/X-ray, Be and B populations}
\label{fig:BeXBeBsepCCD}
\end{figure}

These 3 populations were then separated into their own CCDs so that the 3 distributions could be analysed more easily (Figure \ref{fig:BeXBeBsepCCD}). This showed that although the Be stars had more spread in their (J--K) colours than the Be/X-ray stars, the Be/X-ray stars did, indeed, occupy a region within the main body of the Be type stars. The separation of the B type stars were however clear on the left of the plot but with a possible branch towards the Be and Be/X-ray region.

Due to the similarity between the distributions of the Be and Be/X-ray populations, 2-dimensional Kolmogorov-Smirnov (KS) tests were performed on the populations to determine the probability that they were drawn from the same parent population. The KS test used was the version written by Peter Yoachim\footnote{http://www.astro.washington.edu/users/yoachim}under Interactive Data Language (IDL). The comparison was made between these two distributions drawn from the (J--K) vs. (3.6--4.5$\mu$m) CCD. This returned a probability of 0.0174 that these two distributions were drawn from the same parent population, therefore not reaching the 3$\sigma$ level. A similar test was performed on the Be/X-ray and B populations which returned a probability of $7.4\times10^{-6}$ that they were drawn from the same parent population. So, though there is some suggestion of differences between the Be stars in Be/X-ray binaries and those in isolated systems, it is clear that both those groups are clearly distinguished from the isolated B stars. This is, of course, to be expected if the presence, or otherwise, of a circumstellar disk is the key ingredient in producing the IR excess.In fact, looking again at Figure \ref{fig:BeXBeBsepCCD}, there is a strong suggestion that a significant part of the population of previously-identified B stars are incorrectly labelled as such, and that the Be star fraction should be higher.

\begin{figure}
\centering
\includegraphics[angle = -90,width=90mm]{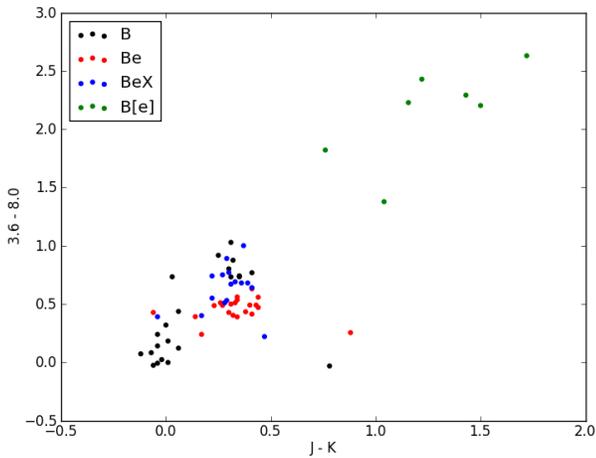}
\caption{Colour-colour diagram of Be/X-ray, Be, B[e] and B populations.}
\label{fig:BeXBeBB[e]CCD}
\end{figure}

Compact accreting objects are present in at least 2 Galactic sgB[e] systems  (CI Cam ~\citep{Bartlett2013} and IGR13618-4848 ~\citep{Chaty2012}) and X-ray detections have also been made of 2 sgB[e] stars in the MCs (Bartlett et al., in prep). Although it is not clear what is the origin of the X-ray emission is in these systems (accretion or colliding winds), it was decided to extend this comparison to include known B[e] stars in the SMC. According to ~\cite{Bonanos2010} there are 7 such stars identified in the Spitzer/SAGE data. Because these stars are characterised by extended IR emission from a cool shell or toroid, the IR colour range was extended to (3.6--8.0$\mu$m). This reduced the size of the sample of B, Be \& Be/X-ray binaries, but left enough to enable a meaningful comparison - see Figure \ref{fig:BeXBeBB[e]CCD}.

The B[e] stars can be seen to occupy the upper right region of this plot. They are quite distinct from the other populations, and hence this CCD confirms that none of the Be/X-ray population included in this work includes a possible B[e] star. Clearly these kind of CCD plots provide good discriminators between the differing B star populations.

\subsection{Outbursts from SXP sources}

The SMC was observed twice using Spitzer's IRAC and MIPS instruments as part of the SAGE programme ~\citep{Gordon2011} during two epochs in June and September 2008({\it epochs 1 \& 2}, respectively). The IRAC catalogue contains observations of objects within the SMC in all 4 of its bands. In addition, this catalogue includes a third epoch, refered to as {\it epoch 0}, which used data taken from a previous study of the SMC performed in May 2005. This enabled the variability of the SXP objects to be studied on the two different timescales, a few months and a $\sim$3 years.

\begin{figure}
\centering
\includegraphics[angle = -90, width=90mm]{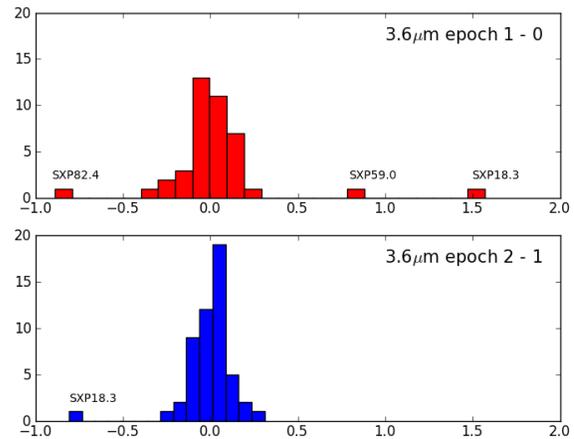}
\caption{Magnitude difference between epochs showing the time variance in the brightness of SXP 18.3, SXP 82.4 and SXP 59.0 in the 3.6$\micron$ band.}
\label{fig:DeltaMagHist}
\end{figure}

Histograms of the change in magnitudes between epochs, for the IRAC bands, were plotted to check for any object showing significant variability (Figure \ref{fig:DeltaMagHist}). The long time scale between epoch 0 and the other two epochs (3 years) revealed that several objects' magnitudes had substantially changed. The most significant case was that of SXP 18.3 (Table \ref{tab:SXP183mag}). The extreme X-ray variability of this object during this period of time has been reported by ~\cite{Schurch2009}.

\begin{figure}
\centering
\includegraphics[angle = -0, width=90mm]{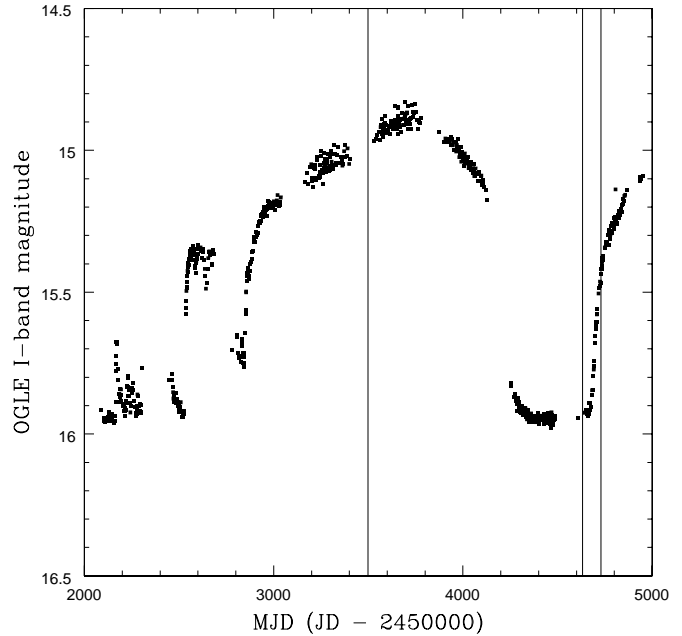}
\caption{The OGLE I-band lightcurve of SXP 18.3. The epoch of the three first Spitzer/IRAC measurements are indicated by vertical lines. Later Spitzer observations are not covered by OGLE observations.}
\label{fig:sxp183phot}
\end{figure}

\begin{table*}
\centering
    \begin{tabular}{ccccccccc}
    \hline
        Epoch   & 3.6$\mu$m mag & Error & 4.5$\mu$m mag & Error & 5.8$\mu$m mag & Error & I-band mag & Error \\ \hline
        Epoch 0 (?? May 2005) & 14.35        & 0.02  & 14.24        & 0.02  & 14.22        & 0.07  & 14.968         & 0.002  \\
        Epoch 1 (15 Jun 2008)& 15.92        & 0.09  & 15.94        & 0.10  & null         & null  & 15.925         & 0.005  \\
        Epoch 2 (19 Sep 2008)& 15.11        & 0.06  & 14.95        & 0.06  & null         & null  & 15.435         & 0.004  \\
        SAGE-Var epoch 1  (17 Aug 2010)& 15.27        & 0.13  & 15.04        & 0.14  & null         & null  & null         & null  \\
        SAGE-Var epoch 2  (12 Sep 2010)& 15.47        & 0.11  & 15.37        & 0.11  & null         & null  & null         & null  \\
        SAGE-Var epoch 3 (24 Dec 2010)& 16.12        & 0.18  & 15.99        & 0.13  & null         & null  & null         & null  \\
        SAGE-Var epoch 4  (16 Jun 2011)& 16.04        & 0.13  & 16.35        & 0.20  & null         & null  & null        & null  \\
\hline
    \end{tabular}
\caption{Spitzer/IRAC magnitudes of SXP 18.3, plus the nearest I-band measurement in time from OGLE.}
\label{tab:SXP183mag}
\end{table*}

SXP 18.3 showed a change of one magnitude in the 4.5 $\micron$ band between epochs 1 and 2, with this large change occuring on a time scale of just 3 months. On a longer timescale changes of over a magnitude are present in both the 3.6 $\micron$ and 4.5 $\micron$ bands. The Optical Gravitational Lensing Experiment project has an SMC X-Ray variables monitoring system (OGLE-XROM)\citep{Udalski2008}, and it obtained frequent I-band (0.8$\micron$) data. These data that showed that SXP 18.3 is a very variable object which revealed strong flares in the I-band close to the time of the SAGE observations - see Figure~\ref{fig:sxp183phot}.

\begin{figure}
\centering
\includegraphics[angle = -90,width=90mm]{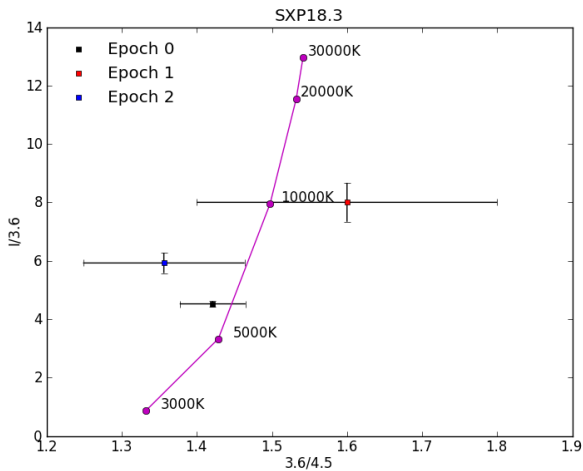}
\caption{SXP 18.3 epoch flux ratio comparison with theoretical blackbody values.The vertical axis shows the ratio of the fluxes in the OGLE I-band (0.8$\micron$) to the IRAC (3.6$\micron$). The horizontal axis shows the flux ratio of the two IRAC bands (3.6$\micron$ and 4.5$\micron$)}
\label{fig:sxp183}
\end{figure}

In order to achieve near-simultaneity between the Spitzer data and the XROM data, the exact dates of the Spitzer individual observation were determined and the closest corresponding I band data for these dates were then found on XROM. OGLE/XROM data only exist for the first three Spitzer observations. This enabled the I band to be included in CCD plots of SXP 18.3 for these first 3 epochs. The resulting CCD showed great variation in this object’s colour - see Figure ~\ref{fig:sxp183}

 The flux values for the 3.6 and 4.5 $\micron$ bands used were those from the SAGE-SMC catalogue ~\citep{Gordon2011} and the OGLE I band magnitude was converted to flux. Flux values for simple blackbodies of different temperatures were then calculated and these values plotted on the same diagram. This diagram indicates the variation in the average temperature of the circumstellar disk component SXP 18.3 over the 3 epochs. This object was observed to be at its brightest during epoch 0 and its faintest during epoch 1. The blackbody curve, however, suggests that SXP 18.3 is at its coolest during epoch 0 and its hottest during epoch 1. Thus this plot implies that the object cools as it gets brighter and the warms up again as the I-band declines. Since the far-IR emission is primarily from the circumstellar disk which is believed to be optically-thick, then the brighter state almost certainly indicates a larger disk size which is, on average, cooler than the smaller, though less-bright disk.

\subsection{H$\alpha$ Equivalent Widths}

\begin{figure}
\centering
\includegraphics[angle = -0, width=80mm]{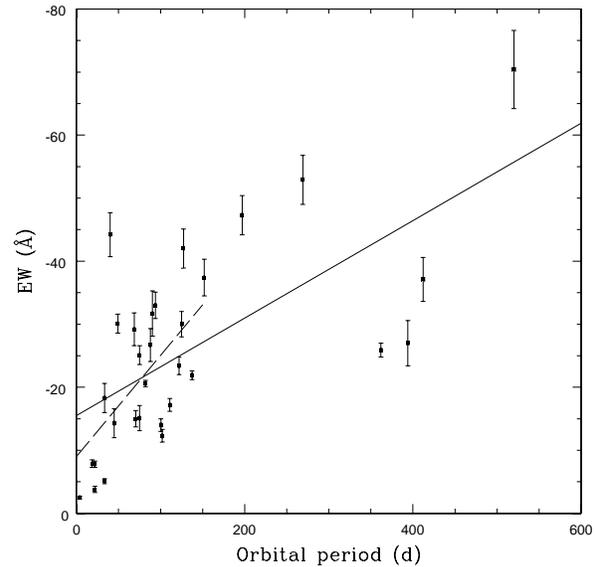}
\caption{Orbital period plotted against H$\alpha$ Equivalent Width (EW) for SXP objects. The linear best fit line to all the data is shown (solid line), and the fit to just the data with $P_{orb} \le 150d$ (dashed line). Errors bars are shown for both parameters, though fractionally very small for the orbital period. }
\label{fig:PorbHa}
\end{figure}

H$\alpha$ is emitted from the circumstellar disc that surrounds the Be stars and is therefore another good probe of the disc. The equivalent width (EW) of the H$\alpha$ emission line was plotted against orbital period for each of the SXP systems with confirmed values - see Figure \ref{fig:PorbHa}. The H$\alpha$ EW and the orbital period show a strong correlation with a Pearson Correlation coefficient of R=0.644 (corresponding to a probability of $9.4\times10^{-5}$ that the correlation is random). This is a much stronger result than previously reported and agrees with that found by ~\cite{Reig1997} for the separate population of Be/X-ray systems in the Milky Way. This correlation is thought to show evidence for truncated circumstellar discs ~\citep{Dachs1986} arising from interaction with the orbiting neutron star. The positive correlation between H$\alpha$ and orbital period suggests that systems with larger orbital periods have larger H$\alpha$ values and therefore can sustain much larger discs. The linear best fit to all the data is shown in Figure \ref{fig:PorbHa} as a solid line and gives the relationship:

\begin{equation}
H\alpha EW(\AA) = [-0.08 \times P_{orb}] - 15.5
\end{equation}

$P_{orb}$ is the orbital period in units of days. There is a strong suggestion that this best fit line is over-estimating the H$\alpha$ EW at lower orbital periods, so a fit to only those points with a $P_{orb} \le 150d$ is also shown as a dashed line. This gives a slightly lower correlation coefficient of R=0.51 for just those points and the relationship:

\begin{equation}
H\alpha EW(\AA) = [-0.16 \times P_{orb}] - 9.7
\end{equation}

\begin{figure}
\centering
\includegraphics[angle = -0, width=80mm]{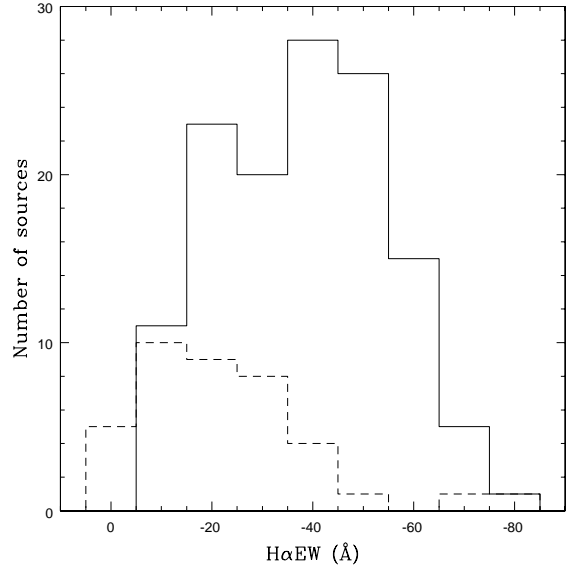}
\caption{Distribution of H$\alpha$ EW values for a sample of single Be stars in the SMC (solid line)~\protect\citep{Martayan2007}, compared to the same parameter for the sample of SXP sources presented in this work (dashed line).}
\label{fig:hacomp}
\end{figure}

~\cite{Reig1997} suggested this relationship arises from the disc of the Be star being truncated by the neutron star during its orbit. This implies that the size of the circumstellar disk in Be/X-ray systems should, on average, be smaller than those in isolated Be stars. To investigate this further the distribution of H$\alpha$ EW values for our SXP sources presented in Table~\ref{tab:smcopt} is shown in Figure~\ref{fig:hacomp}. For comparison, the H$\alpha$ EW values from the catalogue of isolated Be stars in the SMC ~\citep{Martayan2007} is also shown in this figure. The figure strongly suggests that, indeed, the EW values of the SXP population are, on average, lower than those of the isolated Be star population. The mean value of the SXP sample is -27\AA ~and that of the isolated Be stars -40\AA. Furthermore, a  Kolmogorov-Smirnov test of the two populations gives a probability of 0.9\% that they are drawn from the same population.

A more direct test of this proposed physical link between the size of the neutron star orbit, a, and the size of the circumstellar disk, $R_{cs}$, would be to determine both of these parameters for each system and then see if a correlation exists. To determine the size of the circumstellar disk the H$\alpha$EW values presented in Table~\ref{tab:smcopt} were used and inserted into the relationship from \cite{Hanuschik1989}:

\begin{equation}
log(\sqrt(\frac{R_{OB}}{R_{{cs}}})= [-0.32\times log(-EW)]-0.2
\end{equation}

In this expression $R_{OB}$ is the radius of the Be star determined from the individual spectral types given for each source in Table~\ref{tab:smcopt}, and EW is the H$\alpha$ EW values in \AA.

The size of the semi-major axis, a, of the neutron stars orbit may be determined from Kepler's Third Law:

\begin{equation}
\label{eqn:kepler}
a=[\frac{P_{orb}^{2}G(M_{ns}+M_{OB})}{4\pi^{2}}]^{1/3}
\end{equation}

where $M_{OB}$ is the mass of the specific Be star determined from the given spectral type, $M_{ns}$ is the mass of the neutron star (assumed here to be $1.4M_{\odot}$), and $P_{orb}$ is the orbital period.

\begin{figure}
\centering
\includegraphics[angle = -0, width=80mm]{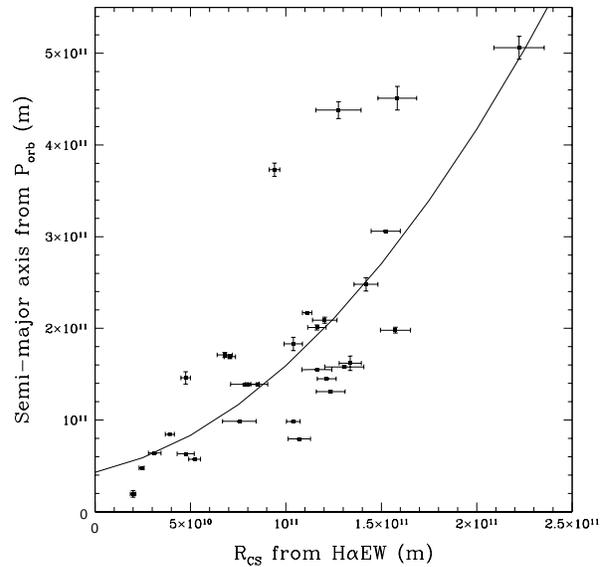}
\caption{This figure shows the size determination in metres of the semi-major axis of the neutron star orbit (y-axis) and the size of the circumstellar disk (x-axis) for a sample of 31 SXP sources. The continuous line is a best fit quadratic function given in Equation~\ref{eqn:fit}.}
\label{fig:size}
\end{figure}

Using those two equations the sizes of the disk and the orbit were calculated for each source and the results are presented in Figure~\ref{fig:size}. It is clear that there is a strong correlation between these two basic physical parameters - the Pearson correlation coefficient, R=0.74. This corresponds to a probability of only $1\times10^{-5}$ that there is no correlation; almost an order of magnitude stronger support for a relationship than is seen in the secondary characteristics plotted in Figure \ref{fig:PorbHa}.

A simple best fit to the data is also shown in Figure~\ref{fig:size}. Both linear and quadratic functions were explored and the best fit was found to be the following expression to determine the size of the orbital semi-major axis in metres, a, as a function of the size of the circumstellar disk, $R_{cs}$, also in metres:

\begin{equation}
\label{eqn:fit}
a = (7\times10^{-12})(R_{cs})^{2} + 0.4524R_{cs} + (4.30\times10^{10})
\end{equation}

This equation implies that the size of the orbit is 1.5--2.0 times that of the circumstellar disk, consistent with the concept that the neutron star is normally orbiting outside the circumstellar disk.

We note in passing that discrepancies between stellar classification and observed luminosities, plus inferred masses \& temperatures, have been reported for HMXB companions by ~\cite{Conti1978} and ~\cite{Kaper2001}. However, changing the Be star mass by 2-3 $M_{\odot}$ in Equation~\ref{eqn:kepler} has less than a 5\% effect on the value for the semi-major axis and is, therefore, unimportant in this context.

It now seems beyond question that disk truncation is occurring in systems with neutron star companions.

\section{Conclusions}

This work presents the first large-scale, coherent sample of Be/X-ray sources in the Small Magellanic Cloud. Since this sample matches, or even exceeds in number the Milky Way sample, it provides an excellent astrophysical test bed for binary evolution and accretion processes. This is made even more effective because the coherent location of all these sources, with a well-defined distance and interstellar absorption, means that many of the issues arising from similar studies of Galactic objects are avoided. Therefore it is expected that this SMC population will provide an extremely useful astrophysical sample for testing future theoretical work.

\section{Acknowledgements}

We are grateful to Christian Knigge for re-running the KMM test on the revised SMC pulsar sample. We are also grateful to David Reibel for providing the results of the Spitzer SAGE-Var observations of SXP 18.3 before they became public.

\bibliography{mjc2015}{}
\bibliographystyle{mn2e}

\newpage

%
\begin{table*}
\setlength\tabcolsep{2.5pt}
\small
\centering
\caption{X-ray properties of the Small Magellanic Cloud sources - Part 1. See text for detailed explanation of the contents.}
\begin{tabular}{rcccccclcccc}
 Short ID&&RA(2000)&&&Dec(2000)&&Alternative name(s)&Pulse&Orbital&Other&Eccen\\
&h&m&s&d&m&s&&(secs)&(days)&(days) &{\it e}\\
 \hline
    SXP 2.16&?&&&?&&&RXTE SMC2165 - sky location unclear&2.16&&&\\
    SXP 2.37&0&54&36.2&-73&40&35.00&SMC X-2&2.37&9.30&&0.07$\pm$0.02\\
    SXP 2.76&0&59&12.80&-71&38&44.00&RX J0059.2-7138&2.76&81.81&&\\
    SXP 3.34&1&5&9.00&-72&11&46.9&AX J0105-722, RX J0105.1-7211&3.34&10.72&1.099& \\
    SXP 4.78&0&52&11.00&-72&20&18.00&XTE J0052-723, - sky location unclear&4.78&&&\\
    SXP 5.05&0&57&2.30&-72&25&55.00&IGR J00569-7226 &5.05&17.20&&0.16$\pm$0.02\\
    SXP 6.62&0&54&46.30&-72&25&23.00&CXOU J005446.2-722523&6.62&&&\\
    SXP 6.85&1&2&53.30&-72&44&35.00&XTE J0103-728, [SHP2000] SMC 100&6.85&22.00&&0.26$\pm$0.03\\
    SXP 6.88&0&54&46.30&-72&25&22.80&[MA93] 798&6.88&2.71&&\\
    SXP 7.78&0&52&5.64&-72&26&04.0&SMC X-3&7.78&44.80&&\\
    SXP 7.92&0&57&58.00&-72&22&29.50&CXOU J005758.4-722229, [SHP2000] SMC 75 &7.92&35.61&&\\
    SXP 8.80&0&51&53.20&-72&31&48.80&RX J0051.8-7231,AzV 102&8.90&33.40&&0.41$\pm$0.04\\
    SXP 9.13&0&49&13.60&-73&11&37.80&AX J0049-732&9.13&40.10&&\\
    SXP 9.60&?&&&?&&&RXTE source sky location unclear&9.60&&&\\
    SXP 11.5&1&4&42.30&-72&54&3.70&IGR J01054-7253, [M2002] SMC 59977&11.48&36.30&&0.28$\pm$0.03\\
    SXP 11.6&1&57&16.15&-72&58&32.60&IGR J015712-7259&11.60&35.40&&\\
    SXP 11.9&0&48&13.47&-73&22&3.10&XMMU J004813.9-732202, M[2002] SMC 10287&11.89&&&\\
    SXP 15.3&0&52&13.98&-73&19&18.80&RX J0052.1-7319,&15.30&74.3&&\\
    SXP 16.6&?&&&?&&&RXTE pulsar - sky location unclear&16.60&33.70&&\\
    SXP 18.3&0&49&11.47&-72&49&37.60&XTE J0055-727, XMMU J004911.4-724939&18.37&17.79&&0.43$\pm$0.03\\
    SXP 22.1&1&17&40.60&-73&30&50.60&RX J0117.6-7330&83.7&&&\\
    SXP 25.5&0&48&14.10&-73&10&4.00&XMMU J004814.1-731003&25.55&22.50&&\\
    SXP 31.0&1&11&8.40&-73&16&46.00&XTE J0111.2-7317&31.00&90.40&&\\
    SXP 46.6&0&53&55.20&-72&26&44.80&1WGA 0053.8-7226,XTE J0053-724&46.60&137.40&&\\
    SXP 51.0&?&&&?&&&RXTE pulsar - sky location unclear(=SXP 25.5?)&51.00&&&\\
    SXP 59.0&0&54&56.20&-72&26&47.90&RX J0054.9-7226,XTE J0055-724&58.95&122.00&&\\
    SXP 65.8&1&7&12.60&-72&35&33.80&CXOU J010712.6-723533&65.80&111.00&&\\
    SXP 74.7&0&49&5.90&-72&50&55.00&RX J0049.1-7250, AX J0049-729&74.80&33.30&2.4 &0.40$\pm$0.23\\
    SXP 82.4&0&52&9.00&-72&38&3.30&XTE J0052-725&82.40&362.00&1.33&\\
    SXP 89.0&?&&&?&&&RXTE pulsar - sky location unclear&89.00&&&\\
    SXP 91.1&0&50&56.90&-72&13&34.40&AX J0051-722 ,RX J0051.3-7216 &91.10&88.00&&\\
    SXP 95.2&?&&&?&&&RXTE pulsar - sky location unclear&95.20&&&\\
    SXP 101&0&57&27.0&-73&25&19.60&AX J0057.4-7325, RX J0057.3-7325&101.40&21.90&&\\
    SXP 138&0&53&23.80&-72&27&15.00&CXOU J005323.8-722715&138.00&125.00&&\\
\end{tabular}
\label{tab:smc1}
\end{table*}

\begin{table*}
\setlength\tabcolsep{2.5pt}
\small
\centering
\caption{X-ray properties of the Small Magellanic Cloud sources - Part 2. See text for detailed explanation of the contents. The last object in this table is technically in the Magellanic Bridge and therefore does not have an SXP nomenclature. No sources in this table have measured eccentricities, so that column is not included here.}
\begin{tabular}{rcccccclccc}
 Short ID&&RA(2000)&&&Dec(2000)&&Alternative name(s)&Pulse&Orbital&Other\\
&h&m&s&d&m&s&&(secs)&(days)&(days) \\
 \hline
    SXP 140&0&56&5.80&-72&21&59.00&XMMU J005605.2-722200,2E0054.4-7237&140.10&197.00&\\
    SXP 144&?&&&?&&&RXTE pulsar - sky location unclear&144.10&59.40&\\
    SXP 152&0&57&50.30&-72&7&56.00&CXOU J005750.3-720756, [MA93]1038&152.10&&\\
    SXP 153&1&7&43.20&-71&59&53.90&XMMU J010743.1-715953&153.00&100.30&\\
    SXP 169&0&52&55.30&-71&58&6.00&XTE J0054-720, AX J0052.9-7158&169.30&68.60&0.755 \\
    SXP 172&0&51&52.00&-73&10&34.00&AX J0051.6-7311, RX J0051.9-7311&172.40&68.8&\\
    SXP 175&1&1&52.50&-72&23&34.00&RX J0101.8-7223 [MA93]1288&175.40&87.2?&\\
    SXP 202A&0&59&21.03&-72&23&17.30&1XMMU J005920.8-722316&202.00&&\\
    SXP 202B&0&59&28.68&-72&37&4.20&XMMU J005929.0-723703&202.00&224.6&\\
    SXP 214&0&50&11.26&-73&0&26.10&XMMU J005011.2-730026&214.00&&\\
    SXP 264&0&47&23.30&-73&12&27.50&XMMUJ004723.7-731226, RXJ0047.3-7312,[MA]172&263.60&49.20&\\
    SXP 265&1&32&51.50&-74&25&45.30&XMMSL1 J013250.6-742544 &264.50&146.00&0.867 \\
    SXP 280&0&57&49.60&-72&2&36.20&AX J0058-72.0&280.40&127.30&\\
    SXP 292&0&50&47.90&-73&18&17.0&CXOU J005047.9-731817&292.00&&\\
    SXP 293&0&58&12.57&-72&30&48.80&RX J0058.2-7231, XTE J0051-727&293.00&59.70&\\
    SXP 304&1&1&2.90&-72&6&59.00&CXOU J010102.7-720658, [MA93]1240,RXJ0101.0-7206&304.50&520.00&0.26 \\
    SXP 323&0&50&44.70&-73&16&5.00&AX J0051-73.3,RXJ0050.7-7316&323.20&&0.71 \\
    SXP 327&0&52&52.20&-72&17&15.10&[SHP 2000] SMC\_45, XMMU J005252.1-721715&327.00&45.99&\\
    SXP 342&0&54&3.90&-72&26&32.90&XMMU J005403.8-722632&342.00&&\\
    SXP 348&1&3&13.97&-72&9&14.80&SAX J0103.2-7209, RX J0103-722&349.90&93.90&\\
    SXP 455&1&1&20.60&-72&11&19.20&RX J0101.3-7211, [MA 93] 1257&452.00&75.00&\\
    SXP 504&0&54&55.90&-72&45&10.90&CXOU J005455.6-724510,RXJ0054.9-7245,AXJ0054.8-7244&503.00&269.00&5.3 \\
    SXP 523&1&2&47.00&-72&4&51.00&	XMMU J010247.5-720450&522.00&&\\
    SXP 565&0&57&36.00&-72&19&34.00&CXOU J005736.2-721934, [MA93]1020&564.80&152.4&\\
    SXP 645&0&55&35.20&-72&29&6.00&XMMU J005535.2-722906&645.00&&\\
    SXP 701&0&55&17.90&-72&38&53.00&XMMU 005517.9-723853&702.00&412.00&0.285 \\
    SXP 723&?&&?&&&&RXTE pulsar - sky location unclear&720.00&&\\
    SXP 726&1&5&55.20&-72&3&51.00&RX J0105.9-7203&726.00&&\\
    SXP 756&0&49&42.00&-73&23&14.70&AX J0049.4-7323, RX J0049.7-7323(?)&755.50&394.00&\\
    SXP 893&0&49&29.85&-73&10&58.30&CXO J004929.7-731058 &892.80&&\\
    SXP 967&1&2&6.70&-71&41&16.20&CXOU J010206.6-714115&967.00&101.4&\\
    SXP 1062&1&27&45.95&-73&32&56.30&2dFS 3831&1063.00&656?&\\
    SXP 1323&1&3&37.40&-72&1&34.10&RX J0103.6-7201, [MA93] 1393, [M2002] SMC 56901&1323.00&&26.2,0.88, 0.41 \\
    SXP 4693&0&54&46.30&-72&25&22.80&CXOU J005446.3-722523 &4693.00&&\\
    0157-7259&1&57&16.15&-72&58&32.6&IGR J015712-7259 Bridge object&11.60&35.1?\\
\end{tabular}
\label{tab:smc2}
\end{table*}

\begin{table*}
\setlength\tabcolsep{3.0pt}
\scriptsize
  \centering
  \caption{Small Magellanic Cloud Be/X-ray systems: optical/IR properties. The JHK values come from the IRSF catalogue of ~\protect\cite{Kato2007} and the longer IR wavelength data from Spitzer ~\protect\citep{Gordon2011}. The H$\alpha$ Equivalant Width (EW) values come from ~\protect\cite{Klus2014}. X-ray sources with no identified optical counterparts are not included in this table.}

\begin{tabular}{ccccccccccccccccccccc}
         ID&Spect&V&Opt col&J&J&H&H&K&K&3.6$\mu$m&3.6$\mu$m& 4.5$\mu$m&4.5$\mu$m&5.8$\mu$m&5.8$\mu$m&8.0$\mu$m&8.0$\mu$m&IR col&H$\alpha$&H$\alpha$\\
&Class&mag&B-V&mag&err&mag&err&mag&err&mag&err&mag&err&mag&err&mag&err&J-K&EW(\AA)&error\\
\hline
    SXP 0.72&BO I&13.15&-0.14&13.62&0.02&13.65&0.01&13.66&0.02&13.38&0.07&13.23&0.04&13.21&0.08&13.01& 0.09&-0.04&-2.5&0.2 \\
    SXP 2.37&O9.5III-V&16.64&0.06&15.1&0.02&15.01&0.02&14.85&0.03&14.29&0.06&14.19&0.05&13.94&0.13&-&-& 0.25&-7.9&0.6 \\
    SXP 2.76&B1IIIe&14.01&0.06&13.96&0.02&13.84&0.02&13.63&0.02&13.37&0.06&13.14&0.04&12.87&0.08&12.71&0.09& 0.33&-20.6&0.5 \\
    SXP 3.34&B1-B2 III-V&15.63&-0.01&15.66&0.01&15.56&0.02&15.45&0.04&14.99&0.07&14.7&0.05&14.29&0.17&-& -&0.21&-85.2&8.9 \\
    SXP 4.78&&&&14.54&0.01&14.58&0.01&14.65&0.02&14.52&0.1&14.71&0.06&14.43&0.16&-&-&-0.11&?-43.7&1.1 \\
    SXP 6.62&&&&15.32&0.01&15.17&0.02&14.92&0.02&14.46&0.06&-&-&13.85&0.12&-&-&0.40&&\\
    SXP 6.85&O9.5-B0 IV-V&14.60&&14.82&0.02&14.74&0.02&14.57&0.02&14.16&0.05&14.1&0.07&14.16&0.16&-&-& 0.25&-3.8&\\
    SXP 7.78&B1-B1.5 IV-V&14.90&0.00&14.8&0.01&14.7&0.01&14.5&0.01&14.09&0.05&13.93&0.06&13.79&0.11&-&-& 0.30&-14.3&2.3 \\
    SXP 8.80&O9.5-B0 IV-V&14.87&-0.27&14.7&0.01&14.64&0.02&14.56&0.02&14.17&0.05&14.07&0.09&13.98&0.12&-& -&0.14&-5.1&0.4 \\
    SXP 9.13&B1-B3 IV-V&16.50&0.10&16.23&0.02&16.02&0.03&15.8&0.04&15.37&0.04&15.08&0.06&-&-&-&-&0.43& -44.2&3.5 \\
    SXP 11.5&O9.5-B0 IV - V&14.80&0.00&14.21&0.08&13.75&0.06&13.74&0.04&-&-&13.3&0.04&13.26&0.09&13.17& 0.1&0.47&&\\
    SXP 11.9&&14.90&&14.67&0.01&14.62&0.01&14.53&0.02&14.42&0.07&14.28&0.07&14.06&0.15&-&-&0.14&&\\
    SXP 15.3&O9.5-B0 III-V&14.70&-0.01&14.4&0.02&14.3&0.01&14.12&0.01&13.6&0.05&13.45&0.05&13.24&0.09& 13.12&0.1&0.28&-25.1&1.5 \\
    SXP 18.3&&16.00&&16.05&0.02&16.03&0.03&16.13&0.05&15.92&0.09&15.94&0.1&-&-&-&-&-0.08&&\\
    SXP 22.1&O9.5-B0 III-V&14.20&-0.04&14.18&0.01&14.11&0.02&13.87&0.03&13.35&0.04&13.23&0.04&13.05&0.08& 12.74&0.07&0.31&-25.1&2.5 \\
    SXP 25.5&B1.5e&15.70&0.01&15.68&0.02&15.68&0.02&15.64&0.04&15.33&0.05&15.17&0.12&-&-&-&-&0.04&&\\
    SXP 31.0&B0.5-B1 V&15.50&-0.10&15.4&0.02&15.3&0.02&15.14&0.03&14.41&0.05&14.14&0.05&-&-&-&-&0.26& -31.7&3.6 \\
    SXP 46.6&O9.5-B1 IV-V&14.70&-0.07&14.72&0.01&14.6&0.02&14.45&0.02&13.62&0.12&13.62&0.1&13.37&0.12&-& -&0.27&-21.9&0.7 \\
    SXP 59.0&O9V&15.30&-0.04&15.52&0.01&15.47&0.02&15.39&0.04&15.32&0.06&15.02&0.08&-&-&-&-&0.13&-23.4& 1.4 \\
    SXP 65.8&B1-B1.5II-III&15.00&&15.64&0.01&15.49&0.02&15.37&0.04&15.03&0.07&-&-&-&-&-&-&0.27&-17.2&\\
    SXP 74.7&B3V&16.90&0.09&16.72&0.03&16.54&0.04&16.32&0.06&15.78&0.11&15.57&0.14&-&-&-&-&0.40&-18.3&2.3 \\
    SXP 82.4&B1-B3III-V&15.00&0.14&15.59&0.01&15.52&0.02&15.47&0.05&14.37&0.06&14.19&0.08&13.8&0.13&-&-& 0.12&-25.9&1.1 \\
    SXP 91.1&B0.5III-V&15.10&-0.08&14.73&0.02&14.63&0.01&14.37&0.02&13.84&0.05&13.67&0.06&13.47&0.09& 13.1&0.08&0.36&-26.7&2.6 \\
    SXP 101&B3-B5 Ib-II&&&15.64&0.01&15.57&0.02&15.38&0.03&15.15&0.06&15&0.08&-&-&-&-&0.26&-7.8&\\
    SXP 138&B1-B2 IV-V&16.20&-0.09&16.1&0.02&15.98&0.03&15.89&0.06&15.43&0.07&15.18&0.08&-&-&-&-&0.21& -30.0&2.0 \\
    SXP 140&B1V&15.90&-0.04&15.34&0.01&15.26&0.02&15.09&0.03&14.95&0.06&14.64&0.07&-&-&-&-&0.25&-47.3&3.1 \\
    SXP 152&B1-B2.5 III-V&15.70&-0.03&15.36&0.01&15.19&0.02&14.98&0.02&14.51&0.06&14.31&0.06&14.24&0.13& -&-&0.38&-17.3&1.7 \\
    SXP 153&B2IV-Ve.&&&16.4&0.02&16.21&0.02&16&0.06&15.13&0.06&14.87&0.09&-&-&-&-&0.40&-14.0&1.0 \\
    SXP 169&B0-B1 III-V&15.50&0.02&15.29&0.01&15.17&0.01&14.93&0.02&14.58&0.06&14.45&0.06&14.07&0.13&-&-& 0.36&-29.2&2.6 \\
    SXP 172&O9.5-B0 V&14.50&-0.07&14.61&0.01&14.53&0.01&14.44&0.02&13.93&0.04&13.72&0.06&13.61&0.08&13.3& 0.07&0.17&-15.0&1.3 \\
    SXP 175&&14.60&-0.04&14.82&0.01&14.73&0.02&14.53&0.02&14.04&0.04&13.71&0.06&13.5&0.07&13.32&0.09& 0.29&&\\
    SXP 202A&B0-B1 V&15.10&-0.24&14.72&0.01&14.65&0.01&14.5&0.02&14.01&0.03&13.74&0.04&13.56&0.07&13.32& 0.07&0.22&-18.1&\\
    SXP 202B&&&&15.26&0.01&15.05&0.01&14.82&0.02&14.38&0.07&14.13&0.07&14.04&0.15&-&-&0.44&&\\
    SXP 214&B2-B3 III&&&15.32&0.02&15.38&0.02&15.37&0.02&15.26&0.05&15.35&0.08&-&-&-&-&-0.05&&\\
    SXP 264&B1-B1.5 V&15.85&0.00&15.92&0.02&15.75&0.02&15.54&0.03&14.92&0.06&14.85&0.08&-&-&-&-&0.38& -30.1&1.7 \\
    SXP 265&B1-2 II-IVe&15.00&&15.92&0.07&16.12&0.08&16.7&0.11&&&&&&&&&-0.78&&\\
    SXP 280&B0-B2 III-V&15.70&-0.12&15.26&0.01&15.05&0.01&14.76&0.02&14.46&0.06&14.2&0.05&14.47&0.16&-&-& 0.50&-42.0&3.1 \\
    SXP 292&O9-B2 III-V&15.1&&15.09&0.01&14.99&0.02&14.85&0.02&14.358&0.05&14.292&0.076&13.863&0.106&&&& -31.0&\\
    SXP 293&B0IIIe&14.90&&14.53&0.01&14.35&0.01&14.14&0.02&13.7&0.04&13.46&0.04&13.23&0.08&13.12&0.1& 0.39&&\\
    SXP 304&B0-B2 III-V&15.70&-0.04&15.56&0.02&15.4&0.02&15.13&0.03&14.69&0.05&14.37&0.07&14.47&0.14&-&-& 0.43&-70.4&6.2 \\
    SXP 323&B0-B0.5 V&15.40&-0.04&15.2&0.02&15.1&0.02&14.9&0.02&14.35&0.05&14.16&0.04&13.87&0.13&13.63& 0.11&0.30&-30.9&1.1 \\
    SXP 327&B2??&&&16.55&0.02&16.27&0.02&16.14&0.04&15.66&0.08&15.65&0.1&-&-&-&-&0.41&&\\
    SXP 342&&&&15.21&0.02&15.17&0.02&15.03&0.03&-&-&14.95&0.08&-&-&-&-&0.18&&\\
    SXP 455&B0.5-B2 III-V&15.50&-0.07&16.03&0.02&15.99&0.02&15.89&0.05&15.55&0.07&15.54&0.07&-&-&-&-& 0.14&-15.1&2.0 \\
    SXP 504&B1 III-V&14.99&-0.02&14.77&0.01&14.63&0.01&14.4&0.01&13.89&0.04&-&-&13.43&0.08&13.07&0.09& 0.37&-52.9&3.9 \\
    SXP 523&&16.00&-0.25&16.58&&16.52&&16.61&&&&&&&&&&-0.03&&\\
    SXP 565&B0-B2 IV-V&15.97&-0.02&15.77&0.01&15.6&0.01&15.35&0.03&14.72&0.07&14.48&0.05&14.28&0.14&-&-& 0.42&-37.4&2.9 \\
    SXP 645&&&&15.11&0.01&15.09&0.01&15.06&0.03&14.1&0.06&13.79&0.05&13.69&0.09&13.29&0.11&0.05&&\\
    SXP 701&O9.5 V&16.00&0.03&15.6&0.02&15.41&0.02&15.21&0.02&14.45&0.07&14.12&0.07&14.03&0.15&-&-&0.39& -37.1&3.5 \\
    SXP 726&&15.60&-0.02&15.47&0.01&15.37&0.02&15.04&0.03&14.56&0.05&14.38&0.05&14.33&0.15&-&-&0.43&&\\
    SXP 756&O9.5-B0.5 III-V&14.98&0.05&14.57&0.01&14.4&0.01&14.16&0.02&13.53&0.04&13.3&0.04&13.03&0.07& 12.95&0.07&0.41&-27.0&3.6 \\
    SXP 893&B1e&&&15.75&0.02&15.46&0.02&15.22&0.02&14.66&0.05&14.44&0.07&14.15&0.1&-&-&0.53&&\\
    SXP 967&B0-B0.5 III-V&14.60&&14.46&0.02&14.39&0.01&14.24&0.02&13.63&0.05&13.41&0.04&13.25&0.08&12.98& 0.09&0.22&-12.3&\\
    SXP 1062&B0-0.5III&&&14.17&0.01&14.07&0.01&13.88&0.02&13.41&0.05&13.2&0.05&12.89&0.08&12.65&0.06& 0.29&&\\
    SXP 1323&B0 III-V&14.60&&14.6&0.02&14.51&0.02&14.34&0.02&14.07&0.05&13.88&0.04&13.71&0.07&-&-&0.26& -17.1&1.5 \\
    SXP 4693&&15.40&0.14&15.32&0.01&15.17&0.02&14.92&0.02&14.46&0.06&-&-&13.85&0.12&-&-&0.40&&\\
    \end{tabular}
  \label{tab:smcopt}
\end{table*}

\end{document}